\documentclass[12pt,a4paper]{article}

\usepackage{graphicx}

\begin{document}

\title{Light emission of very low density hydrogen excited by an extremely hot light source;
\\applications in astrophysics.}

\author{Jacques Moret-Bailly.
\\Physique, Universit\'e de Bourgogne, F-21000 Dijon France.}
\maketitle
\begin{abstract}
Str\"omgren studied the action of an extremely hot source on a diluted pure hydrogen cloud; a very ionized, spherical hydrogen plasma surrounded by neutral atomic hydrogen is formed. A relatively thin intermediate, partially ionized, hydrogen shell, is cooled by the radiation of the atoms.

Str\"omgren was unaware of that this plasma, similar to the plasma of a gas laser, can be superradiant at several eigen frequencies of atomic hydrogen; the superradiant rays emitted tangentially with the sphere appear resulting from a discontinuous ring because of the competition of optical modes. The superradiance intensely depopulates the excited levels, including the continuum of proton-electron collisions, by cascades of transitions combined into resonant multiphotonic transitions so that the gas is cooled brutally beyond the radius of the Str\"omgren sphere.

The extreme brightness of the rays emitted by the source allows a multiphotonic non-resonant absorption leading in stationary states or the ionization continuum. This absorption combines with the superradiant emissions in a multiphotonic diffusion induced by the superradiant rays. Although its brightness remains higher than that of the superradiant rays, the source becomes invisible if it is observed through a small solid angle.

The lines emitted inside the sphere are all the more weak as they arrive of an internal area, lower in atoms, and more reddened also by a parametric transfer of energy towards the thermal radiation catalyzed by excited atomic hydrogen present in the sphere only.

The Str\"omgren sphere appears to help to simply explain the appearance and the spectrum of supernova 1987A.

PACS {42.65.Es, 42.50.Md, 95.30.Jx}
\end{abstract}

\section{Introduction}

Astrophysicist Str\"omgren \cite{Stromgren} studied the structure and the spectroscopy of a model with spherical symmetry made up of a very hot source (like an O or B star) surrounded by hydrogen. Later work on this model, founded on the assumption of a local reabsorption of the lines of neutral hydrogen atoms (assumption \textquotedblleft one the spot\textquotedblright), or criticizing this assumption \cite{Ritzerveld}, did not modify qualitatively the properties of the model.

\medskip

Section 2 applies to Str\"omgren model the theories of coherent optics and spectroscopy, old, but which were really used only with the development of the microwave and laser sources, so that the word \textquotedblleft coherent\textquotedblright {} is often replaced by \textquotedblleft microwave\textquotedblright {} or \textquotedblleft laser\textquotedblright.

So certain terms like \textquotedblleft star\textquotedblright, \textquotedblleft Str\"omgren sphere\textquotedblright {} make think of astrophysics, it acts, in this section, only of practical abbreviations to represent simple physical systems, \textquotedblleft star\textquotedblright {} substitute, for example \textquotedblleft very hot source of light\textquotedblright.

The study of this section requires a good knowledge of the theories founding coherent spectroscopy (Huygens, Rayleigh, \dots) and more recent works (Planck \cite{Planck11}, Einstein \cite{Einstein17}) which are at the base of the development of the applications of microwaves and lasers optics and spectroscopy, such as the multiplications, combinations, shifts of frequencies.

Section 3 compares the appearances and the spectra obtained in section 2 with astronomical images and spectra, relating to particularly supernova 1987A.

So that the results of section 2 are usable in section 3 without a true reading, each result obtained in section 2 is explained there in italic, at the end of the sub-sections.

\section {Optics and spectroscopy of a Str\"omgren sphere.}
\subsection {General reminder.}
The model of Str\"omgren is stationary, made of:

- a relatively small very hot ($T>10^6$ K) source $O$ of light able to emit light in far ultraviolet, up to X rays.

- a huge cloud of hydrogen, supposed cold at its outside.

A spherical symmetry is assumed around the center of the source $O$. 

The models of Str\"omgren sphere studied until now are founded on individual interactions of the atoms with the electromagnetic field of great brightness emitted by the source.

But Einstein \cite{Einstein17} showed that absorption and emission of a spectral line of frequency $\nu$ can be studied simultaneously:
the variation $\mathrm dL$ of spectral brightness\footnote{Except contrary specification the brightnesses considered in this text are spectral brightnesses, and not the integrals of spectral brightnesses extended to all frequencies.} $L$ of a ray on an infinitesimal way $\mathrm dx$ is $\mathrm dL = aL(N_2-N_1)\mathrm dx$, where $N_2$ and $N_1$ are the populations of molecules (mono- or poly- atomic), respectively in the upper or lower levels of the corresponding transition. 

Boltzmann law defines the radiant temperature $T_{1,2}$ of a medium, at the frequency $\nu$ by $N_2/N_1= exp(-w/kT_{1,2})$, with $w=h\nu$; a negative value of $T_{1,2}$ given by a inversion of population is accepted by convention.
Planck law \cite{Planck11,Einstein13,Nernst} defines the temperature $T_\nu$ of a ray from its brightness: $L=\frac{2h\nu^3}{c^2} (\frac{1}{exp(h\nu/kT_\nu)-1}+\frac{1}{2})$. In a black body, $T_\nu = T_{1,2}$; in an opaque medium at the frequency $\nu$, these temperatures are equalized except if a contribution of energy maintains imbalance.

The amplification of rays of light by application of Einstein law \cite{Einstein17} does not change the geometry of the luminous rays, therefore that of wave surfaces. It is said that the amplification is coherent; this coherence which results from the thermodynamic calculation of Einstein, will be explained more physically in subsection \ref{refrac} on the example of the refraction.

Let us notice that in a black body at $0 K$, brightness is nonnull; but it cannot provide an usable energy because the minimal energy would have to be reduced; the field which corresponds to it is stochastic and its amplification provides the spontaneous emission.
Also, so that the coherent amplification is observed without being drowned in the stochastic, spontaneous emission, it is necessary that the brightness of the rays to be amplified, is definitely higher than ${h\nu^3}/{c^2}$. For this reason, coherent amplifications are spectacularly observable only with bright sources, in microwaves, now with lasers. The rays emitted by hot stars have a large brightness too.

\medskip

If an elementary volume of source is crossed by a particularly luminous ray, according to Einstein law, this ray extracts much more energy than quite other ray. The supplied energy being generally limited, the other rays are weakened. This \textquotedblleft competition of the modes\textquotedblright {} lets remain with a notable brightness only the most brilliant, \textquotedblleft superradiant\textquotedblright {} rays; thus, a laser emits only some brilliant rays.

Masers and lasers are sets of a superradiant medium and a resonant cavity. Without resonant cavity, the spontaneous emission, temporally incoherent is amplified; on the contrary, the amplification of the field preserved by the resonant cavity gives a temporally coherent emission. Some superradiant sources observed in astrophysics, wrongly, were qualified maser: they are not temporally coherent for lack of resonant cavities in space.

\medskip
When an atom is strongly excited, it can lose its energy by a cascade of successive jumps, emitting several lines successively. If the emission of these lines is made very fast, for example because they are induced, the emissions become simultaneous, forming a non-linear interaction named \textquotedblleft multiphotonic interaction\textquotedblright.

\medskip
Superradiance and multiphotonic interactions can reduce the \textquotedblleft frictions\textquotedblright {} which are opposed to a fast increase in the entropy of a system of light and matter.

\medskip
{\it Conclusion: The coherent, induced emission of the light discovered by Einstein, supports, by competition of the modes in a strongly emitting medium, the appearance of an intense superradiance in some modes only; this superradiance at the eigen frequencies of the matter, causes a rapid increase in the entropy.}

\subsection{Reminder on coherent light-matter interactions; example: refraction}\label{refrac}
To explain the wave propagation, Huygens deduced from a wave surface known at time $t$, a later wave surface, at time $t+\Delta t$. For that, he supposes that each element of volume contained between wave surfaces relating to times $t-\mathrm dt$ and $t$ emits at the local speed of the waves $c$, a spherical wavelet of radius $c\Delta t$; the set of these wavelets has as envelopes the sought wave surface and a retrograde wave surface canceled by the retrograde waves emitted at other times.

Let us suppose that each element of volume considered by Huygens emits also a wavelet of much lower amplitude whose phase is delayed by $\pi/2$. The interaction of light and matter can be modeled classically by a non resonant molecular oscillator, or by a quantum \textquotedblleft dressing\textquotedblright {} of the matter whose stationary state is mixed with other states. Except for the second order, this emission does not modify the initial amplitude. Are $E_0 \sin (\Omega T)$ the incident field, $E_0K\epsilon \cos(\Omega T)$ the field diffused in the layer of thickness $\epsilon=c\mathrm dt$, and $K$ a coefficient of diffusion; the total field is:

\begin{equation}
E=E_0[\sin(\Omega t)+K\epsilon \cos(\Omega t)]\label{refr}
\end{equation}
\begin{equation}
\approx E_0[\sin(\Omega t)\cos(K\epsilon)+\sin(K\epsilon )\cos(\Omega t)]=E_0\sin(\Omega t -K\epsilon).
\end{equation}
 This result defines the index of refraction $n$ by the identification 
\begin{equation}
K=2\pi n/\lambda=\Omega n/c.\label{indice}
\end{equation}

When a refracted wave is delayed by $\pi/2$ by the refraction, one can estimate that the total amplitude was completely absorbed then re-emitted with a delay of $\pi/2$. The necessary path $e$ checks the equation $Ke-2\pi e/\lambda=\pi/2=2\pi(n-1)e/\lambda$; for $\lambda=.5\mu m$, and $n=1.33$, $e=0.75\mu m$; in a swimming pool, one still clearly sees an object 75 meters away, the ratio of the diffused intensities coherent/incoherent is thus of the order of 10$^8$.

The large value of this ratio is not due to the molecular properties, but to the addition of amplitudes or intensities, plus the weakness of the sources of incoherence which are: i) fluctuations of density; ii) resonant absorptions followed by spontaneous emissions, reduced, out of resonance, to diffusions induced by the stochastic field (phosphorescences). In diluted gases, the fluctuations of density correspond to the collisions; the density of binary collisions is a quadratic function of the pressure.

\medskip

{\it Conclusion: Among the interactions of the light and matter, the refraction is the best known example of parametric interactions (parametric means coherent without permanent excitation of the matter); this example shows that the coherent effects are much more intense than the incoherent effects.}

\subsection{The Str\"omgren shell.}\label{shell}
Between the Str\"omgren sphere containing a hydrogen plasma almost entirely ionized (dissociated in protons and electrons) and an area containing non excited atoms primarily, is a plasma containing very excited neutral atoms; this medium is similar to plasma which form the active matter of the gas lasers excited by an electric current. Str\"omgren showed that this layer is relatively thin by taking account only of its spontaneous radiation at the eigen frequencies of hydrogen. By taking account of its induced radiation, its thickness is still reduced. 

Let us consider a ray passing at a distance $r$ of the source. If it does not cross the Str\"omgren shell, it is not amplified, it remains dark.
If it passes close to the source, it crosses the same mediums as a more distant ray, but under a lower incidence, therefore on shorter paths. 
There thus exists at least a distance $R$ for which amplification is maximum and the ray is superradiant. 

The superradiance appears at the distance R tangentially with Str\"omgren sphere; indeed, if the superradiant ray moved towards the interior of the sphere, it would be necessary to re-examine the choice of R; the tangential rays with the sphere cross the same areas as the rays which move away from there more quickly; they thus are more amplified and gain the competition of the modes. 

\begin{figure}
\includegraphics[height=6 cm]{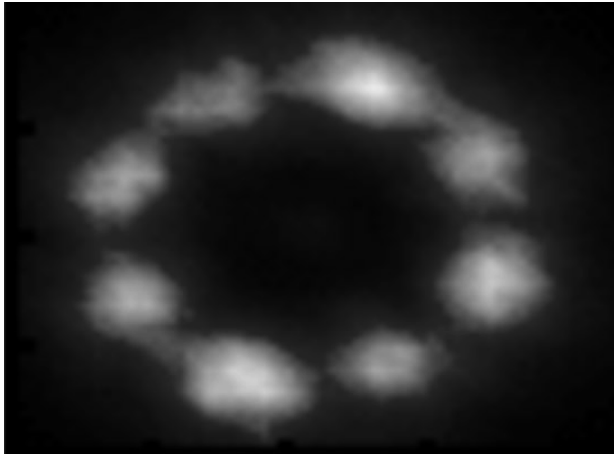}
\caption{\label{laser} (Modes of a laser. The number of spots may be larger; their shapes may differ. From: J. M. POL, Thesis, University of the Balearic Islands, 2002.}
\end{figure}

Thus, for an observation into a given direction, the superradiant rays are placed on the generators of a cylinder tangent at the sphere; the competition of the modes in this cylinder lets see only spots similar to those which are observed with certain lasers. The laser figure \ref{laser} shows such a system of spots; the number of spots can be much larger and their form different, so that one obtains, for example with \textquotedblleft daisy modes\textquotedblright, a figure similar to the petals of a daisy.

For a given transition, the superradiance tends to quickly balance, for each transition, the radiant temperature of the medium and the temperature of the rays; to amplify a line, the high level HI of a transition depopulates for the benefit of the low level LO; this supports the appearance of news superradiances populating the level HI or depopulating LO, which has two consequences:

- i) The superradiant rays tend to have the same geometry at several frequencies, although their invariants of Clausius, about the square wavelength $\lambda^2$, are different: the spots of diffraction of the shorter wavelength rays will tend to be placed as close as possible of the sphere, while remaining in the spots corresponding to the largest wavelengths. Thus \textquotedblleft polychromatic modes\textquotedblright {} are defined.

- ii) The de-energizing of a very high level by a cascade of transitions binds these transitions in a process of multiphotonic emission.

\medskip

\begin{figure}
\includegraphics[height=6 cm]{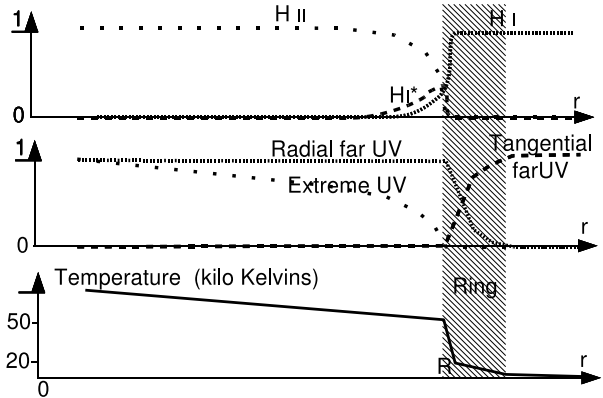}
\caption{\label{rad} Variation of the relative densities of H$_I$, H$_{II}$ and excited atomic hydrogen H$_I$*, relative intensities of light, and temperature along a radius.}
\end{figure}

The superradiances intensely de-energize all excited levels, so that the layer containing these levels is much thiner than according to the initial model of Str\"omgren; the speed of de-energizing is limited only by the time of generation of collisions, that the superradiances make pass from elastic to inelastic. Consequently, the collisions generate directly non excited atoms by multiphotonic radiation (Figure \ref{rad}).
Thus, we will pose that the external limit of the Str\"omgren shell does not correspond to the disappearance of excited atoms, but with the disappearance of an amplification of the superradiant rays.

{\it Conclusion: The Str\"omgren shell is a relatively thin spherical zone of transition between the Str\"omgren sphere almost entirely ionized and an area of non excited neutral atoms. Superradiants rays at various eigen frequencies of atomic hydrogen, and at the very low frequencies of the continuum are emitted tangentially with the Str\"omgren sphere and are observed, in a given direction, because of the competition of the modes, like spots forming a circular ring, more brilliant on the internal edge of this ring.

The superradiance brutally de-energizes the excited hydrogen atoms, of which the density was just enough to start tangentially the superradiance on the sphere of Str\"omgren. The electron-proton collisions become inelastic by yielding by superradiant multiphotonic transitions, the energy which leads to non excited atoms.}

\subsection{Parametric interactions in the Str\"omgren shell.}
As the radial rays are very bright, having initially a temperature about $10^6 K$, they are absorbed by atomic hydrogen (excited or not), with any wavelength, in a multiphotonic process which leads in a state dissociated or not. It was seen, in sub-section \ref{shell} that these states are quickly de-energized because of superradiance.

These high energy processes are connected so quickly that their stages are bound into a parametric process, i.e. coherent, and which does not modify finally the state of the atoms: During the process, the atoms are \textquotedblleft dressed\textquotedblright {} by the fields, in a nonstationary state, combination of the initial stationary state with many excited states. The excited levels of the hydrogen atom remain virtual.

Because of the multiphotonic intensity of the process, the temperature of the rays emitted by the star tends towards a common temperature of the superradiant rays: the spectrum of the radial rays tends towards the spectrum of the black body at the temperature of the superradiant rays; as the superradiant lines are sharp, the total brightness (integrated on a broad spectral band) of the radial emission remains definitely larger than the total brightness of the superradiant rays. But, if one supposes that the solid angle of observation of star is much smaller than the solid angle of observation of the superradiants modes, only the latter are visible.

{\it Conclusion. The high temperature of the radiation resulting from star, and the average temperature of the superradiant radiations authorize coherent multiphotonic interactions known as \textquotedblleft parametric\textquotedblright {} because the matter is not excited in a permanent way. Thus, a great proportion of the luminous radiation of the star (continuous spectrum) is transferred to the superradiant radiation having the spectrum of atomic hydrogen. Let us set that the external limit of the Str\"omgren shell corresponds to a stop of the increase in the entropy of the set of the radial and superradiant rays.}

\subsection{Light emitted inside Str\"omgren sphere.}\label{para}
The plasma of protons and electrons is transparent; it appears atoms only beyond a radius for which the temperature is about 50.000 K. These atoms, formed excited or excited by the radiation, emit spontaneously their spectral lines; the emission of these lines by incoherent scattering is negligible if the pressure is low, therefore if the fluctuations of density by collisions are rare. The emission increases quickly according to the distance from the star because the density of neutral atoms increases quickly when the temperature drops. The intensity emitted in the surface layers of the sphere is thus higher than the intensity emitted in the deep layers.

This light emitted mainly in the external layers of the shell is subjected to a coherent Rayleigh diffusion which refracts it, which has little consequence.

Can there be also an observable coherent Raman scattering?

A coherent Raman scattering generates wave surfaces identical to the incident ones, exactly as a Rayleigh scattering does. But the problem of the interference of the scattered beam with the incident one is very different because in a Raman scattering, the difference of phases between these beams at a scattering atom, equal to zero at the beginning of an exciting pulse, increases linearly with time unless a collision destroys this behavior. Thus, beats generally appear, and an elementary spectroscopy is able to split the incident and scattered frequencies. However, if the duration of the exciting light pulse is shorter than the period of the beats and than the collisional time, it will be shown that the interference produces mainly a single frequency, the remainder being eliminated by destructive interferences dues to the scattering on different wave surfaces. This is a particular case of a condition of space coherence and constructive interference which is, according to G. L. Lamb Jr \cite{Lamb}: the pulses must be \textquotedblleft shorter than all relevant time constants\textquotedblright. Setting $\omega$ the Raman frequency, $K'>0$ the anti-Stokes diffusion coefficient, formula \ref{refr} becomes:
\begin{equation}
E=E_0[(1-K'\epsilon)\sin(\Omega t)+K'\epsilon \sin((\Omega+\omega)t)].
\end{equation}
In this equation, incident amplitude is reduced to obtain the balance of energy for $\omega=0$.
\begin{equation}
E=E_0\{(1-K'\epsilon)\sin(\Omega t)+K'\epsilon[\sin(\Omega t)\cos(\omega t)+\sin(\omega t)\cos(\Omega t)]\}.
\end{equation}
$K'\epsilon$ is infinitesimal; suppose that between the beginning of a pulse at $t=0$ and its end, $\omega t$ is small; the second term cancels with the third, and the last one transforms:
\begin{eqnarray}
E\approx E_0[\sin\Omega t+\sin(K'\epsilon\omega t)\cos(\Omega t)]\nonumber\\
E\approx E_0[\sin(\Omega t)\cos(K'\epsilon\omega t)+ \sin(K'\epsilon\omega t)\cos(\Omega t)=E_0\sin[(\Omega+K'\epsilon\omega)t].\label{eq4}
\end{eqnarray}
Hypothesis $\omega t$ small requires that Raman period $2\pi/\omega$ is large in comparison with the duration of the light pulses; this is a first Lamb's condition; the second is that collisional time must be larger than this duration.

Stokes contribution, obtained replacing $K'$ by a negative $K\textquotedblright$, must be added. Assuming that the gas is at equilibrium at temperature $T$, $K'+K\textquotedblright$ is proportional to the difference of populations in Raman levels, that is to $\exp[-h\omega/(2\pi kT)]-1 \propto \omega/T$.

$K'$ and $K\textquotedblright$ obey a relation similar to relation \ref{indice}, where Raman polarisability which replaces the index of refraction is also proportional to the pressure of the gas $P$ and does not depend much on the frequency if the atoms are far from resonances; thus, $K'$ and $K\textquotedblright$ are proportional to $P\Omega$, and $(K'+K\textquotedblright)$ to $P\Omega \omega/T$. Therefore, for a given medium, the frequency shift is:

\begin{equation}
\Delta\Omega=(K'+K\textquotedblright)\epsilon\omega\propto P\epsilon\Omega\omega^2/T.\label{delom}
\label{redshift}
\end{equation}

The relative frequency shift $\Delta\Omega/\Omega$ of this space-Coherent Raman Effect on time-Incoherent Light (CREIL) is nearly independent on $\Omega$ \cite{Mor98b,Moret}.

For the largest shift with Lamb's conditions, the limit value of $P$ and $\omega$ are inversly proportional to $\tau$, so that the order of magnitude of the shift is, from equation \ref{delom} proportional to $\tau^{-3}$. Replacing the femtosecond laser pulses by the nanosecond incoherent light pulses multiplies $\tau$ by $10^5$, the CREIL effect by $10^{-15}$ : The CREIL cannot be observed on natural light in an Earth size laboratory! But, as CREIL shifts add along large paths in excited hydrogen, it may be observed in space.

There remains a problem: The atoms dressed by the electromagnetic fields must return to their eigenstate after a light pulse; as usual in coherent spectroscopy (frequency doubling, mixing...) the solution is an interaction between several waves, with a zero balance of energy for the atoms. To increase entropy, the hot beams, often light, must loose energy through a decrease of frequency while the coldest, usually radio-waves, in particular thermal background, are heated by increases of frequencies.

Usual incoherent light may be modeled by pulses of some nanoseconds, so that applying Lamb's conditions, the pressure of gas must be low and Raman frequency must be lower than 1GHz.

As the shift $\Delta\Omega$ is proportional to $\omega^2$, a strong shift requires a Raman frequency as large as allowed by first Lamb's condition.
With atomic hydrogen, frequency 1420 MHz of the hyperfine transition in 1S state is too large; in the first excited state the frequencies 178 MHz in the 2S$_{1/2}$ state, 59 MHz in 2P$_{1/2}$ state, and 24 MHz in 2P$_{3/2}$ are very convenient.

\medskip

$K'$ and $K\textquotedblright$ are increasing functions of strong fields as $K$ and the indices of refraction, so that using powerful femtosecond laser pulses, the CREIL interactions increase, making laboratory experiments easy (Yan et al. \cite{Yan}, Dhar et al. \cite{Dhar} \dots ). The CREIL becomes the \textquotedblleft Impulsive Stimulated Raman Scattering\textquotedblright {}  (ISRS).

\medskip

The incoherent radiations, in particular Lyman alpha, emitted in the Str\"omgren sphere are all the more intense as the zone of emission is close to this surface; the lowering of frequency of the lines by transfer of energy to the thermal radiation is all the more intense as the source is deeper. Thus the spectral lines are spread out towards the low frequencies whereas their intensity decrease.

In the inner rim of the shell, where the superradiant beams are born, there is some hydrogen in states 2S and 2P. The beams which interact by a CREIL effect are, by decreasing temperature:

- The radial beams;

- The superradiant beams at optical frequencies;

- The superradiant thermal radiation emitted by the initial decay of the very excited and ionization states;

- The spontaneously emitted beams.

The interaction is weak because the path is short; but a blueshift of the coldest, spontaneously emitted beams could be seen.

In the remainder of the shell, the atoms are not excited, there is no frequency shift. The connexion between frequency shifts and a path in {\bf excited} atomic hydrogen characterizes a CREIL effect.

\medskip

{\it Conclusion. In the external layers of the Str\"omgren sphere, relatively rich in excited hydrogen atoms 2S or 2P, a parametric set of impulsive Raman effects transfers energy from the rays of light (whose frequency is lowered) towards the thermal radiation. Thus, the lines emitted spontaneously in the Str\"omgren sphere are weaker and more shifted towards low frequencies if they are deeper emitted.

In the internal rim of the shell, the obtained spectrum may be slightly blueshifted by a CREIL transfer of energy from the radial and superradiant beams.
}

\begin{figure}
\includegraphics[height=7 cm]{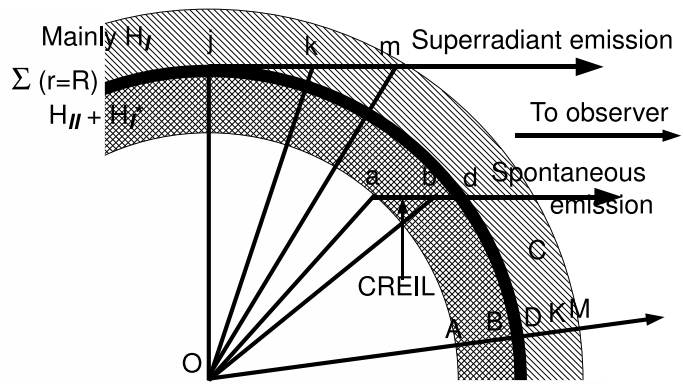}
\caption{\label{diffu}Paths of luminous energy to Earth. On an arbitrary ray, the interactions of light with H I at points A, B, D (=J), K, M are identical to the interactions at points a, b, d, j, k, m placed at same distances of O. Spontaneous emissions following absorptions at a or b (stronger at b) are redshifted by CREIL along ad or bd (larger along ad). Along D K M, the extremely bright radial beam is scattered in a multiphotonic process to amplify the superradiant emission; excited states are so depopulated by superradiance in C region that CREIL is negligible along jkm, so that the Lyman line remains sharp.}
\end{figure}

\subsection{Spectroscopy of other molecules.}

We suppose that other mono- or poly-atomic molecules are mixed with hydrogen, but do not perturb notably the previous study.
Thus a Str\"omgren sphere is generated; out of this sphere hydrogen is mainly atomic until molecules appear under 10 000 K.

Lines emitted spontaneously inside the sphere are strongly redshifted; having a low intensity, they cannot be seen. As the shell is thin, the spontaneous emissions in the shell are not observable.

Out of Str\"omgren shell, the superradiant beams make columns of light, in particular Lyman, able to ionize or simply excite various molecules which make columns of excited gas. As the gas is cold, the lines are sharp; superradiances of these molecules, colinear with the superradiances of hydrogen may appear.

\section{Possible applications to astrophysics.} \label{applic} 

\subsection{Supernova remnant 1987A.} \label{snr1987a}

Str\"omgren's model, improved by holding account of the coherence of the interactions of the electromagnetic waves with hydrogen appears to apply to the actual state of SNR1987A, taking into account the greatest complexity of the hydrogen cloud which surrounds the supernova; in particular, it could help understanding the complexity of the spectra and respond, at least partly, to Heng et al. in their wish that: \textquotedblleft\dots a detailed explanation will require theoretical modeling \dots\textquotedblright {}  \cite{Heng}. 

\medskip

 The similarities are:

The rings appear in a region where temperature is around 50000 K , where density of H$_I$ and H$_{II}$ is low ($10^{10}m^{-3}$), but the available paths, of the order of 0.1 light-year, are widely sufficient for an optical thickness at Lyman lines. Its rings show bright spots corresponding to optical modes, so beautiful that the main, bright ring is sometimes named "pearl necklace"; this main ring shows brighter inner modes (Lawrence et al. \cite{Lawrence}).

The supernova disappeared when the main ring appeared, showing a nearly full transfer of energy from the star to the rings.

All thermal emissions arise from the main ring (Bouchet et al. \cite{Bouchet}).
A large redshift is observed inside the rings, even in soft X rays (Park et al. \cite{Park}).

\medskip

The problems are:

Our model is extremely simplified, its spectroscopy is limited to lines of H$_I$; but Lyman $\alpha$ line plays a particularly important role due to its enormous intensity and its strong interaction with H$_I$ (Pun et al. \cite{Pun}). 

The remnant does not have a spherical symmetry; this may be the result of a non uniform repartition of hydrogen (Sugerman et al. \cite{Sugerman}).

It is difficult to explain the existence of the outer rings (Burrows et al. \cite{Burrows}); maybe the surfaces on which they appear are H$_I$ shells around Str\"omgren spheres generated by the extreme UV radiation of two accreting neutron stars ejected during a first, old explosion of the supernova which produced the axisymmetric structure they observed. Crotts \& Heathcote \cite{Crotts} criticize an hypothesis close to our, which says that a ring corresponds to a limb-brightened ellipsoid; for the outer rings, Sugerman et al. \cite{Sugerman} found by photon echoes two lumbers, on which they drew the emitting rings; slightly moving the weakest parts of the lumbers to obtain ovoids, the rings are where the direction of Earth is tangent to the ovoids. Martin \& Arnett \cite{Martin} show surfaces which could replace a Str\"omgren shell, but the rays which form the principal ring are not tangent; without making the assumption of an axial symmetry, a deformed sand hourglass could solve the problem.

\medskip

A measure of the distance of SN 1987A may use two methods generally considered as reliable:

- When the magnitude of a star varies quickly, it is possible to observe the delay of propagation of light through a direct observation of the star, and the observation of an object enlightened by the star (photon echoes), so that the complete position of the object can be found. Panagia et al. \cite{Panagia} measured, by photon echoes, the absolute radius of the main ring; a division by the angular radius gives a distance of 168000 light-years.

\begin{figure}
\includegraphics[height=6 cm]{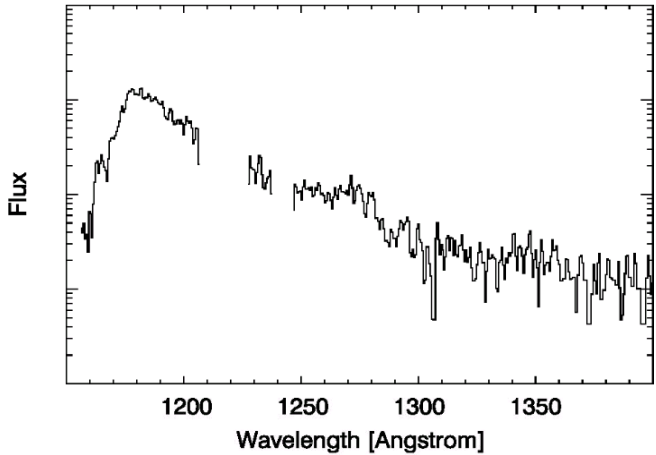}
\caption{\label{mic} Spectrum of SN1987A inside the main ring, from Michael et al. \cite{Michael}.}
\end{figure}

- The spectra observed on and inside the main ring show large red- or blue-shifts whose interpretation by jets of gas is uneasy (Sonneborn et al. \cite{Sonneborn}). In particular, a spectrum (figure \ref{mic}) observed inside the main ring of SN 1987A and published by Eli Michael and 20 co-authors \cite{Michael} is mainly a Lyman forest. Applying Hubble's law shows a distance larger than 2.10$^9$ light-years. As the shape of the source of the spectrum is wide, limited by the main ring, the origin of the spectrum cannot be a punctual quasar beyond the supernova or gas jets. Michael et al. reject {\it implicitly} the Hubble interpretation using numerical computations set on hypothesis of incoherent scatterings similar to the scatterings leading to the \textquotedblleft Wolf effect\textquotedblright {} \cite{Wolf}; the results are similar, not convincing because, for a large redshift, the absorption is too high \cite{James}. The CREIL effect works better (no absorption, unlimited, additive redshifts), and, as observed, works only in the sphere where there is some excited atomic hydrogen.

\subsection{Other applications}\label{misc}

Pioneer 10 and 11 probes show \textquotedblleft anomalous accelerations\textquotedblright {}  when they reach, beyond 5 UA, a region where the solar wind starts to condensate into excited atomic hydrogen (Anderson et al. \cite{Anderson}). A transfer of energy by CREIL effect from solar light to radio-waves catalyzed by 2S or 2P hydrogen produces an increase of their frequencies usually assigned to a Doppler shift due to an anomalous acceleration. A similar amplification of microwave background may explain that some space harmonics seem bound to the ecliptic, that is to the Sun through anisotropies of its wind (Schwarz et al.\cite{Schwarz}).
.

The microwave thermal radiation is anomalously hot in the surroundings of many redshifted bright objects, or hydrogen clouds; for instance,: Croft et al. observed a strong relation between the mean effective optical depth over Ly$_\alpha$ forests and the CMB temperature \cite{Croft}; Verschuur \cite{Verschuur} found an association of diffuse interstellar neutral H$_I$ structure with the brightest peaks in the WMAP ILC map; the origin of these heatings may be a transfer of energy from light redshifted in 2S or 2P atomic hydrogen.

Objects having very different redshifts seem correlated, so that several authors \cite{Arp,Napier,Harutyunian} write that their redshifts are anomalous and search explanations. As these effects are observed for very hot objects, or close to them, excited atomic hydrogen seems involved, a strong argument for a CREIL effect.

Some bright arcs of circle are observed in the sky. Their explanation by an improbable alignment of objects and gravitational lensing, or by shock waves is much more complex than the simply founded, ordinary spectroscopy developed here.

\section{Conclusion.}
Except in laser and microwave technologies, or in refraction, coherent light-matter interactions seem negligible. However, the quasi absence of collisions in a low density gas, forbids the generation of stochastic phase shifts needed, in particular, for incoherent scatterings; on the contrary, in this gas, coherent interactions are not perturbed by collisions.

Induced emissions, multiphotonic interactions and impulsive Raman scatterings allowed us to build a complex physico-optical system from the simple hypothesis of an extremely hot source in a huge cloud of low density hydrogen.

The image of this system is very close to the present image of supernova remnant 1987A. Other not easily understandable astrophysical observation could probably be explained using coherent spectroscopy: laser spectroscopists and astrophysicists would have a lot to gain from a closer cooperation. 

Unhappily, the introduction of apparently new coherent optics sets problems, supporting, for instance, the implicit denial of the validity of Hubble' law expressed by Eli Michael et al. \cite{Michael}, in spectra containing Lyman forests. It may be considered as an attack against the main pillar of the Big Bang theory.


\begin{thebibliography}{}
\bibitem{Stromgren}B. Str\"omgren, ApJ, {\bf 89}, 526 (1939).
\bibitem{Ritzerveld}J. Ritzerveld, A \& A, {\bf 439}, 23 (2005).
\bibitem{Planck11}M. Planck, Verh. Deutsch. Phys. Ges, {\bf 13}, 138 (1911).
\bibitem{Einstein17}A. Einstein, Phys. Zs. {\bf 18}, 121 (1917).
\bibitem{Einstein13}A. Einstein, \& O. Stern, Ann. Physik, {\bf 40}, 551 (1913).
\bibitem{Nernst}W. Nernst, Verh. Deutsch. Phys. Ges {\bf 18}, 83 (1916).
\bibitem{Lamb}G. L. Lamb Jr., Rev. Mod. Phys., {\bf 43}, 99 (1971).
\bibitem{Mor98b}J. Moret-Bailly, Quant. \& Semiclas. Opt., {\bf 10}, L35 (1998).
\bibitem{Moret}J. Moret-Bailly, IEEETPS, {\bf 31}, 1215 (2003).
\bibitem{Yan}Y.-X. Yan , E. B. Gamble Jr. \& K. A. Nelson, J. Chem Phys., {bf 83}, 5391 (1985).
\bibitem{Dhar}Dhar, L. , J. A. Rogers, \& K. A. Nelson, 1994 Chem. Rev. {\bf 94}, 157 
\bibitem{Heng}K. Heng, R. McCray, S. A. Zhekov et al., arxiv:astro-ph/0603151 (2006).
\bibitem{Lawrence}S. S. Lawrence, B. E. Sugerman, P. Bouchet, A. P. S. Crotts, R. Uglesich \& S. Heathcote, ApJ, {bf 537}, L123 (2000).
\bibitem{Bouchet}P. Bouchet, E. Dwek, J. Danziger et al., arxiv:astro-ph/0601495 (2006).
\bibitem{Park}S. Park, S. A. Zhekov \& R. McCray, arxiv:astro-ph/0604201 (2006). 
\bibitem{Pun}C. S. J. Pun ,E. Michael,S. A. Zhekov et al. ApJ, {\bf 572}, 906 (2002).
\bibitem{Sugerman}B. E. K. Sugerman, A. P. S. Crotts, W. E. Kunkel, S. R. Heathcote, S. S. Lawrence, arxiv:astro-ph/0502268 \& 0502378 (2005).
\bibitem{Burrows}C. J. Burrows, J. Krist, J. J. Hester et al., ApJ, {\bf 452}, 680 (1995).
\bibitem{Crotts}A. P. S. Crotts \& S.R. Heathcote, Nature, {bf 350}, 683 (1991).
\bibitem{Martin}C. L. Martin \& D. Arnett, ApJ, {bf 447}, 378 (1995).
\bibitem{Panagia}N. Panagia, R. Gilmozz1, F. Macchetto, H.-M. Adorf \& R. P. Kirshner, ApJ, {\bf 380}, L23 (1991).
\bibitem{Sonneborn}G. Sonneborn, C. S. J. Pun, R. A. Kimble, T. R. Gull, P. Lundqvist et al., Astrophys. J., {\bf 492}, L139 (1998).
\bibitem{Michael}E. Michael, R. McCray, R. Chevalier et al., Astrophys. J., {\bf 593}, 809 (2003).
\bibitem{Wolf}E. Wolf, Nature, {\bf 326}, 363 (1987) 
\bibitem{James}D. James, Pure Appl. Opt. {\bf 7}, 959 (1998)
\bibitem{Anderson}J. D. Anderson, P. A. Laing, E. L. Lau, A. S. Liu, M. M. Nieto \& S. G. Turyshev, ArXiv:gr-qc/0104064 (2002).
\bibitem{Schwarz}D. J. Schwarz, G. D. Starkman, D. Huterer, C. J. Copi, Arxiv:Astro-ph/0403353 (2004)
\bibitem{Croft}R. A. C. Croft, A. J. Banday \& L. Hernquist, Mon. Not. R. Astron. Soc., {\bf 369}, 1090 (2006). 
\bibitem{Verschuur}G. L. Verschuur, arxiv:0704.1125 (2007).
\bibitem{Arp}H. Arp, E. M. Burbidge, Y. Chu, X. Zhu, ApJ {\bf 553}, L11 (2001)
\bibitem{Napier}W. M. Napier, Astrophysics and Space Science, {\bf 285}, 419 (2003).
\bibitem{Harutyunian}H. A. Harutyunian, Astrophysics {\bf 47}, 538 (2005)
\end{thebibliography}
\end{document}